\documentclass{article}
\usepackage{graphicx}
\usepackage{amsmath}
\usepackage{amssymb}

\newcommand{\BiSe}{Bi$_2$Se$_3$}

\title{High pressure transport properties of the topological insulator \BiSe}

\author{J.~J.~Hamlin,$^1$ J.~R.~Jeffries,$^2$ N.~P.~Butch,$^3$ P.~Syers,$^3$ D.~A.~Zocco,$^1$\\ \vspace{0.1in}
S.~T.~Weir,$^2$ Y.~K.~Vohra,$^4$ J.~Paglione,$^3$ and M.~B.~Maple$^1$\\ 
$^1$Department of Physics, University of California, San Diego,\\ \vspace{0.1in}
La Jolla, California 92093, USA\\
$^2$Lawrence Livermore National Laboratory,\\  \vspace{0.1in}
Livermore, CA, 94551\\
$^3$Department of Physics, University of Maryland,\\ \vspace{0.1in}
College Park, MD 20742\\ 
$^4$Department of Physics, University of Alabama at Birmingham,\\
Birmingham, AL, 35294}

\begin{document}

\maketitle

\begin{abstract}
We report x-ray diffraction, electrical resistivity, and magnetoresistance measurements on \BiSe\ under high pressure and low temperature conditions.  Pressure induces profound changes in both the room temperature value of the electrical resistivity as well as the temperature dependence of the resistivity.  Initially, pressure drives \BiSe\ towards increasingly insulating behavior and then, at higher pressures, the sample appears to enter a fully metallic state coincident with a change in the crystal structure.  Within the low pressure phase, \BiSe\ exhibits an unusual field dependence of the transverse magnetoresistance $\Delta \rho _{xx}$ that is positive at low fields and becomes negative at higher fields.  Our results demonstrate that pressures below 8 GPa provide a non-chemical means to controllably reduce the bulk conductivity of \BiSe.
\end{abstract}

\newpage
The material \BiSe\ has been the subject of extensive study for a number of decades, primarily in regard to thermoelectric properties.  In 2009, reports of a topologically protected surface state in \BiSe~\cite{xia_2009_1,zhang_2009_1} lead to an explosion of renewed interest in this material.  The predicted topological surface state is clearly visible in angle resolved photo-emission (ARPES) measurements~\cite{xia_2009_1}.  However, in transport measurements, it has proven difficult to disentangle the surface contribution from the bulk conductivity, which is substantial even in the lowest carrier density samples~\cite{butch_2010_1}.  That the more than 1100 papers concerning \BiSe\ published prior to 2009~\cite{scifinder} failed to bring to light an anomalous surface state suggests that signatures of this surface state in basic measurements such as bulk electrical transport are likely to be subtle.

Several proposed applications for topological insulators~\cite{fu_2008_1,hasan_2010_1} involve the interface between a topological insulator and a superconductor.  Pressure induced superconductivity with $T_c \leq 3$ K has been reported in topological insulator Bi$_2$Te$_3$ and calculations suggest that the surface state may remain intact in the presence of bulk superconductivity~\cite{zhang_2011}. Intercalation of copper between the layers of \BiSe\ has been shown to produce bulk superconductivity with a maximum critical temperature near 4 K~\cite{hor_2010_1}.  Motivation for a high pressure study of \BiSe\ is provided by two possibilities: 1.\ pressure could suppress the bulk conductivity, thereby helping to unmask the surface contribution, or 2.\ pressure might drive \BiSe\ superconducting without destroying the surface state, offering a means to explore the topological insulator/superconductor interface.  In this context, pressure is a particularly powerful tool because (at least in the hydrostatic limit) it provides a means to tune the electronic structure without introducing the disorder inherent to chemical substitution studies.

While there have been studies of the structure and electrical resistivity of \BiSe\ under pressure at temperatures above room temperature~\cite{vereshchagin_1965_1,atabaeva_1973_1}, the electrical resistivity at low temperature has yet to be reported.  We therefore set out to measure the transport properties of \BiSe\ at high pressure and low temperature.  Since pressure was applied at room temperature in all of our experiments, we also performed high pressure x-ray diffraction measurements to determine the structural evolution at room temperature.

All measurements were carried out on high mobility, nominally stoichiometric single crystals of \BiSe, grown as previously reported~\cite{butch_2010_1}.  Electrical transport and optical measurements confirm that these materials are n-type doped semiconductors~\cite{butch_2010_1,sushkov_2010,jenkins_2010,cho_2011}.  Typical values for the carrier density and mobility for this batch of samples were $2 \times 10^{17}$/cm$^3$ and 15,000 cm$^2$/Vs respectively.  Room temperature x-ray diffraction measurements were performed in a diamond anvil cell at sector 16BM-D at Argonne National Laboratory's Advanced Photon Source synchrotron facility.  The wavelength of the x-rays was 0.4134 \AA{}.  Powdered \BiSe\ and copper (as manometer) were loaded into a Ni gasket along with neon as a pressure transmitting medium.  The two dimensional image plate patterns were converted to one dimensional $2\Theta$ versus intensity data using the Fit2D software package~\cite{fit2d}.  The resulting diffraction patterns were fit via Rietveld refinement using GSAS+EXPGUI~\cite{GSAS}.

High pressure electrical transport measurements were carried out using piston-cylinder, Bridgman anvil, and diamond anvil cells.  The diamond cell utilized a ``designer diamond anvil''~\cite{weir_2000_1,jackson_2006} containing eight symmetrically arranged, deposited tungsten micro-probes encapsulated in high-quality homo-epitaxial diamond.  The distance between voltage leads was $\sim 30$ $\mu$m.  The piston-cylinder cell utilized a nearly hydrostatic 1:1 mixture of n-pentane and isoamyl alcohol as the pressure transmitting medium, while the Bridgman and diamond anvil cells used quasi-hydrostatic solid steatite.  Pressure was determined using either the superconducting transition of a chip of Sn~\cite{smith_1969_1} or Pb~\cite{wittig_1988_1} located near the sample (piston cylinder and Bridgman anvil cell, respectively) or ruby fluorescence spectrum~\cite{mao_1986_1} (diamond anvil cell).  In all cases, pressure was adjusted at room temperature.  The resistance of each sample was measured in the \textit{ab}-plane using a four wire configuration and a Linear Research Inc.\ LR-700 AC resistance bridge.  The piston-cylinder and Bridgman cell data extend from room temperature down to $\sim 2-3$ K, while the diamond cell data extends down to 15 K.  Due to uncertainties in the geometry of the small samples and placement of the leads (which may move during pressurization), the absolute resistivity values for the Bridgman cell experiment are accurate only to within a factor of $\sim 2$.
\begin{figure}
  \begin{center}
    \includegraphics[width=0.8\textwidth]{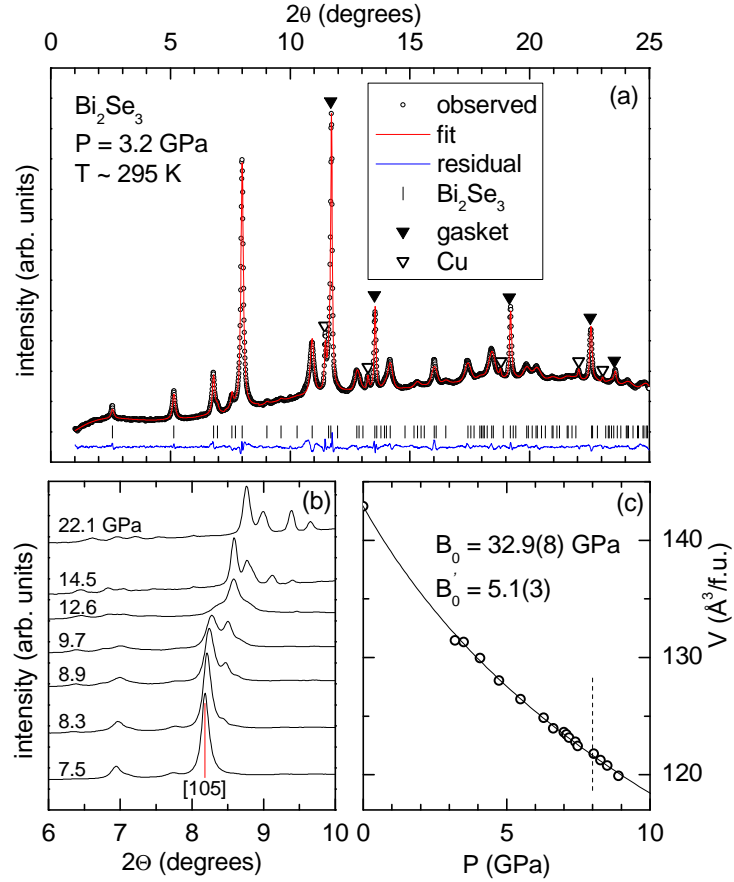}
  \end{center}
  \caption{(a) Diffraction data at 3.2 GPa fit using the rhombohedral (R$\overline{3}$m) Bi$_2$Te$_3$-type structure.  Filled and open triangles indicate reflections due to the gasket and copper manometer respectively. (b) Evolution of the diffraction pattern with pressure.  A structural transition, beginning above $\sim 8$ GPa, is signaled by the growth of several new reflections.  (c) Equation of state of \BiSe\ within the pressure range where the diffraction data can be fit by the R$\overline{3}$m structure.  The solid line is a fit to the data using a third order Birch-Murnhagan equation of state. The dashed vertical line indicates the approximate onset of the structural transition.}
  \label{fig:fig1}
\end{figure}

At pressures below $\sim 8$ GPa, the x-ray diffraction patterns show that \BiSe\ remains in the rhombohedral (R$\overline{3}$m) Bi$_2$Te$_3$-type structure.  An example of a Rietveld refinement of the x-ray diffraction data at 3.2 GPa is shown in Figure~\ref{fig:fig1}a.  The pattern is well fit by a combination of rhombohedral \BiSe, the Cu pressure marker, and the Ni gasket.  Upon increasing pressure at room temperature, a structural phase transition begins to occur near 8 GPa, as illustrated in Figure~\ref{fig:fig1}b by the suppression of the [105] peak of rhombohedral \BiSe\ and the development of several new peaks.  The diffraction pattern at high pressure does not appear to match the structures previously proposed for compressed \BiSe\ (at elevated temperatures), although the appearance of additional peaks suggests that the transition is likely to a structure of lower symmetry.  Further efforts are currently underway to identify the structure of the high pressure phase.  Figure~\ref{fig:fig1}c shows the equation of state of \BiSe\ within the low pressure structure.  The data have been fit using a third order Birch-Murnhagan equation of state~\cite{birch_1947_1}, resulting in a bulk modulus $B_0 = 32.9(8)$ GPa and derivative $B_0^{\prime} = 5.1(3)$.  The bulk modulus is similar to that of the analogous sulfur compound Bi$_2$S$_3$ (37 GPa)~\cite{lundegaard_2005}.
\begin{figure}
  \begin{center}
    \includegraphics[width=0.8\textwidth]{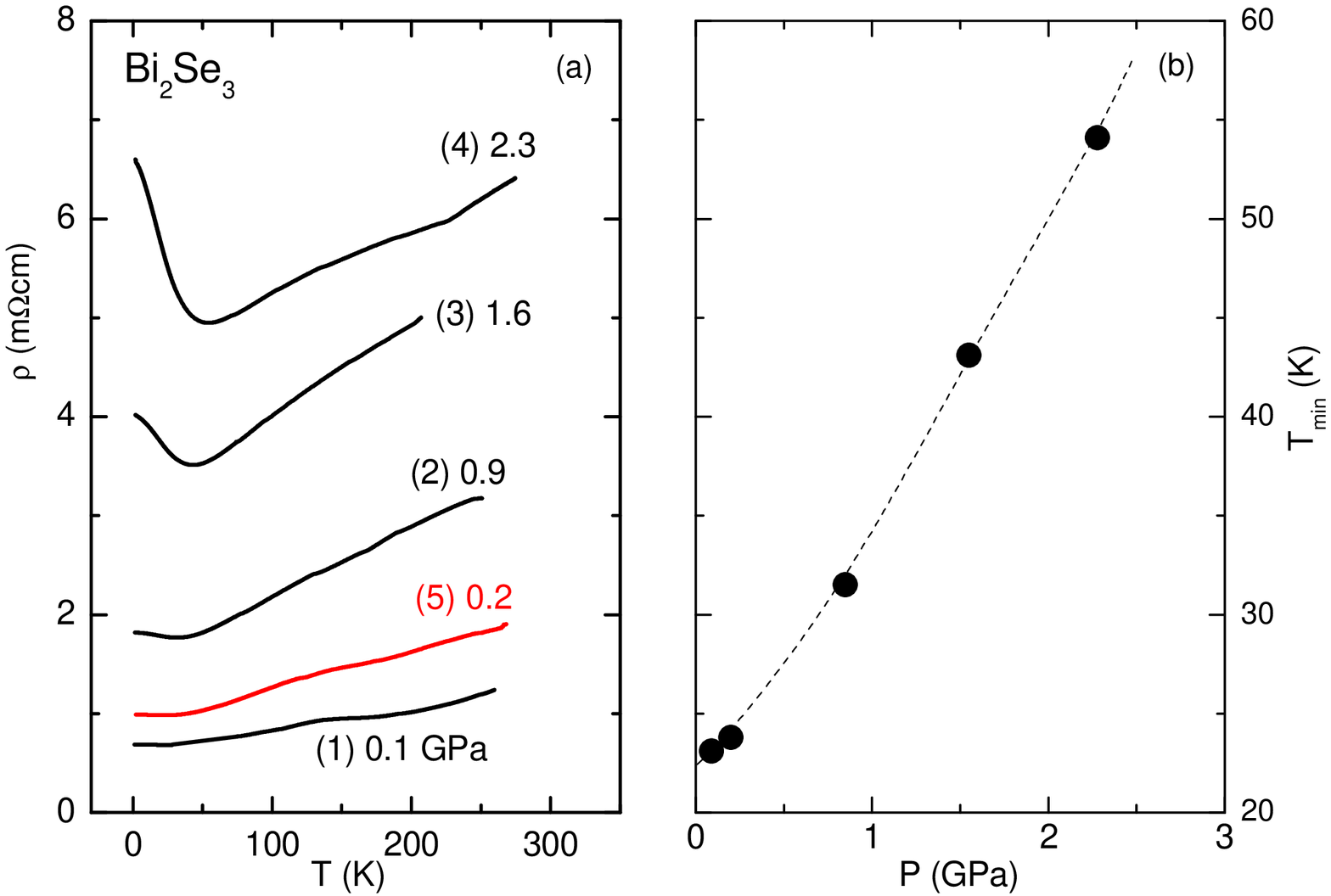}
  \end{center}
  \caption{(a) Electrical resistivity of \BiSe\ versus temperature measured in the hydrostatic cell.  The numbers in parentheses give the order of the measurements (b) Pressure dependence of the electrical resistivity minimum.}
  \label{fig:fig2}
\end{figure}

Figure~\ref{fig:fig2}a illustrates the temperature dependence of the resistivity for the nearly hydrostatic piston-cylinder cell measurements.  Measurements were performed at 0.1, 0.9, 1.6, 2.3, and 0.2 GPa, in that order.  Increasing pressure leads to an increase in the resistivity over the entire measured temperature range.  A curve measured on unloading the pressure (red line in Figure~\ref{fig:fig2}a) shows that the overall resistivity is reversible with pressure.  The low temperature upturn in the resistivity, which is barely visible near ambient pressure, is significantly strengthened as pressure increases.  In addition, $T_{min}$, the temperature at which the resistivity minimum occurs, increases with pressure from $\sim 25$ K near ambient pressure to $\sim 54$ K at 2.3 GPa (Figure~\ref{fig:fig2}b).  At the lowest temperatures, the upturn in the resistivity  shows a tendency to saturate.

There appears to be no clear consensus on the origin of the resistivity minimum. Among samples of widely varying selenium vacancy concentration, $T_{min}$ always appears near 30 K, though the upturn is barely visible in high carrier density samples and becomes significantly larger in low carrier density samples~\cite{butch_2010_1}.  One possibility is that $T_{min}$ represents a crossover from degenerate semiconducting behavior above $T_{min}$ to thermally activated conduction below $T_{min}$ with pressure driving the donor impurity energy level away from the bottom of the conduction band so that de-ionization occurs at progressively higher temperatures.  However, Hall measurements at ambient pressure do not indicate a substantial reduction in the carrier density below 20 K~\cite{hor_2009}, which appears at odds with the above picture.  Resistance minima have also been observed in thin films of \BiSe, with $T_{min}$ varying from $\sim 25$ K to 5 K in 2 to 6 quintuple layer thick samples and the resistance diverging logarithmically at low temperatures~\cite{liu_2011}.  Those authors attributed the increase in resistance at low temperature to electron-electron interactions in the presence of disorder.  Similar resistivity minima near 30 K were also reported for nano-ribbons of \BiSe~\cite{cha_2010}.  The probable presence of small amounts of Fe dopants in the nano-ribbons lead the authors to ascribe the low temperature upturn to a Kondo effect.  Though it is possible that the resistive minima in bulk, thin film, and nano-ribbons of \BiSe\ all derive from different effects, the fact that they occur in a similar range of temperatures is suggestive of a common origin.
\begin{figure}
  \begin{center}
    \includegraphics[width=0.8\textwidth]{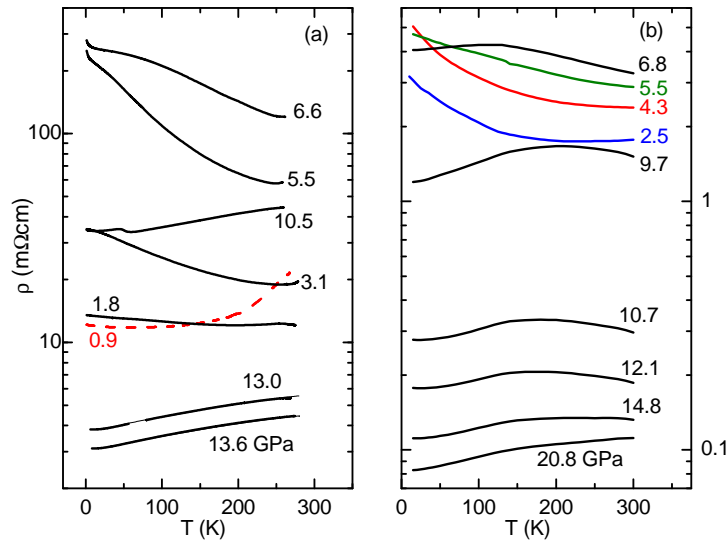}
  \end{center}
  \caption{Electrical resistivity measured in the less hydrostatic Bridgman cell (a), and diamond anvil cell (b).  Data were taken upon monotonically increasing the pressure.}
  \label{fig:fig3}
\end{figure}

\begin{figure}
  \begin{center}
    \includegraphics[width=0.8\textwidth]{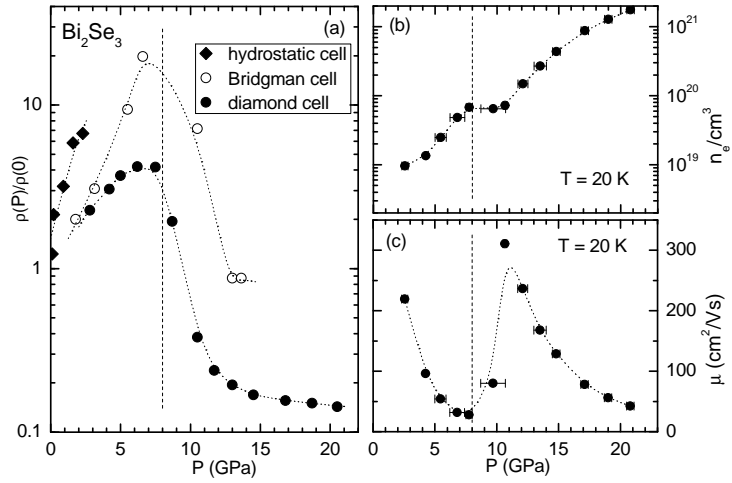}
  \end{center}
  \caption{(a) Normalized room temperature electrical resistivity versus pressure for all three experiments, showing an initial increase in the resistivity with pressure, followed by roughly an order of magnitude drop as the crystal structure changes above $\sim 8$ GPa (dashed vertical line).  The dotted curves are guides to the eye. (b) Carrier density and (c) mobility at 20 K extracted from measurements in the designer diamond anvil cell.  Over the measured pressure range, the carrier density increases by more than two orders of magnitude.  Within the low pressure R$\overline{3}$m phase, the mobility decreases monotonically.  As the crystal structure changes above $\sim 8$ GPa, the mobility increases substantially.}
  \label{fig:fig4}
\end{figure}

Figures~\ref{fig:fig3}a and \ref{fig:fig3}b present the electrical resistivity versus temperature at several pressures for the less hydrostatic Bridgman and diamond anvil cell experiments.  The measurements were taken while monotonically increasing the pressure.  The resistivity is plotted on a logarithmic scale in order to facilitate comparison between the various curves.  Figure~\ref{fig:fig4}a shows a comparison of the pressure dependence of the resistivity for the measurements in the piston-cylinder, Bridgman anvil, and diamond anvil cell experiments.  For all three types of measurements, the room temperature resistivity initially increases with pressure.  As pressure increases above the 8 GPa structural transition, both the Bridgman cell and diamond cell experiments show a large drop in the resistance.

The details of the resistive behavior vary somewhat between the three types of measurements.  For example, the room temperature resistance near 2 GPa scatters over nearly an order of magnitude for the various experiments.  This is not at all surprising given that the resistivity and carrier density of \BiSe\ at ambient pressure can vary by several orders of magnitude depending on the selenium vacancy concentration~\cite{butch_2010_1}.  Differences in the initial carrier concentration and varying degrees of strain could also be the origin of the somewhat different temperature dependences of the resistivity for the three measurements.  Nonetheless, the broad trends for the three experiments are in agreement, with pressure initially driving \BiSe\ towards increasingly insulating behavior before the resistivity drops substantially over the entire temperature range as the pressure is increased above the $\sim 8$ GPa structural transition.

\begin{figure}
  \begin{center}
    \includegraphics[width=0.8\textwidth]{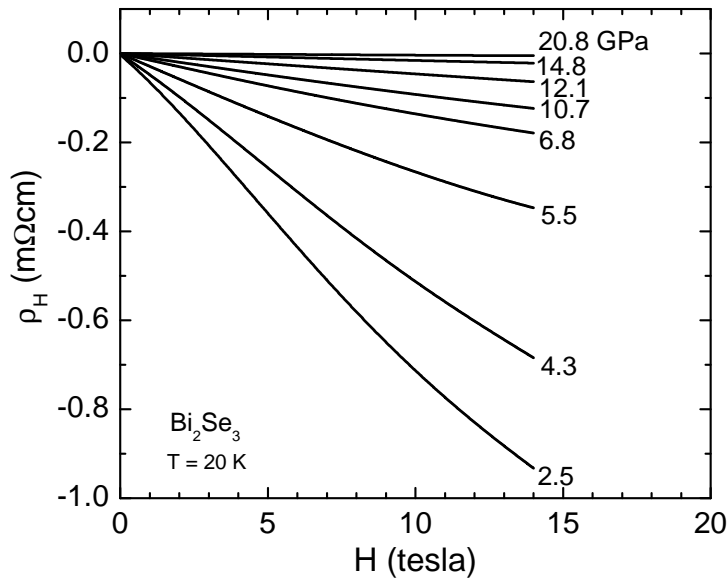}
  \end{center}
  \caption{Symmetrized Hall resistivity data at 20 K from designer diamond anvil cell measurements.}
  \label{fig:fig5}
\end{figure}

At a fixed temperature of 20 K and magnetic fields up to 14 tesla, we carried out Hall and transverse magnetoresistance measurements at several pressures in the designer diamond anvil cell.  Figure~\ref{fig:fig5} shows the field dependence of the Hall resistivity, $\rho _H$, at several pressures.  Values of the Hall coefficient $R_H = \rho _H/H$ were determined from linear fits to the low field portion of symmetrized data.  The carrier density, according to the relation $n_H = (R_He)^{-1}$ is plotted in Figure~\ref{fig:fig4}b.  Assuming that this simple estimate remains valid, \BiSe\ is $n$-type over the entire measured pressure range and, from 2 to 20 GPa, the carrier density increases by more than two orders of magnitude.  From the electrical resistivity $\rho$ and Hall coefficient $R_H$, we estimate the carrier mobility $\mu = R_H/\rho$, which is plotted in Figure~\ref{fig:fig4}c.  Within the low pressure structural phase ($P \lesssim 8$ GPa), the mobility decreases with pressure before increasing sharply as the crystal structure transforms to the metallic high pressure phase.  Within the high pressure phase, the mobility again begins to decrease.  The Hall data indicates that the increase in resistivity is not a simple matter of depleting carriers with pressure.  Rather, the increase in resistivity is connected to a large decrease in the mobility.  One possible explanation for this decrease is that a change in the electronic structure leads bulk carriers to be transfered from the conduction band minimum at the $\Gamma$ point to another extremum with reduced mobility (see, \textit{e.\ g.,} Paul~\cite{paul_1961}).

\begin{figure}
  \begin{center}
    \includegraphics[width=0.8\textwidth]{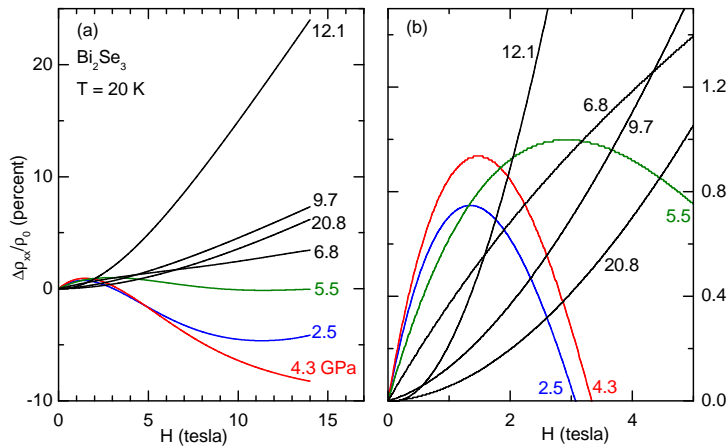}
  \end{center}
  \caption{Transverse magnetoresistance $\Delta \rho_{xx} / \rho _0$ vs $H$ at several pressures. Figure~\ref{fig:fig6}(a) shows the data over the entire field range, while Figure~\ref{fig:fig6}(b) highlights the behavior at lower fields.}
  \label{fig:fig6}
\end{figure}

Figure~\ref{fig:fig6} presents transverse (field perpendicular to current) magnetoresistance data.  At pressures above the structural/insulator-metal transition, the magnetoresistance at low fields is positive and consistent with typical $\Delta \rho _{xx} \propto \mu H^2$ behavior (see Figure~\ref{fig:fig6}b).  At low pressures, the magnetoresistance is quite unusual given that \BiSe\ is described as a single-band material.  At low pressures, the magnetoresistance first increases with field, then decreases and becomes negative at higher fields, before curving back upwards at the highest fields.  Many doped semiconductors display a magnetoresistance that is negative at low fields before crossing zero and becoming positive at higher fields~\cite{alexander_1968,roth_1963}.  This behavior has been explained in terms of a superposition between a weak localization (WL) effect that dominates at low fields 
and a conventional positive $H^2$ contribution that dominates at higher fields~\cite{kawabata_1980}.  In contrast to the latter behavior, at pressures between 2.5 and 6.8 GPa, \BiSe\ exhibits a positive contribution to the magnetoresistance that dominates at the lowest fields.  The magnitude of this positive contribution appears too high to be explained by surface 2D weak anti-localization (WAL) effects that have been observed in thin films of \BiSe~\cite{kim_2011,kim_2011_arxiv}.  A maximum in the magnetoresistance also appears to have been observed in samples of \BiSe\ that were calcium doped to push the chemical potential into the bulk band gap~\cite{checkelsky_2009}, as well as in Cr-doped thin films~\cite{zhang_2011_crossover}.  In addition, rather similar transverse magnetoresistance curves have been observed in strained HgTe~\cite{germanenko_1994}.

In summary, pressure initially drives \BiSe\ towards increasingly semiconducting behavior.  Above 8 GPa, the material undergoes a combined insulator-metal/structural phase transition.  The enhancement of the electrical resistivity in the low pressure phase suggests that pressure could help to unmask the surface state by suppressing the bulk conductivity.  However, Hall data, collected using a designer diamond anvil, shows that this suppression of conductivity occurs through a sharp decrease in mobility and that \BiSe\ remains $n$-type, with a carrier density that increases over the entire measured pressure range.  The substantial increase in carrier density with pressure might be expected to further obscure the surface conduction.  An interesting question is whether the topological surface state might be implicated in the low temperature upturn in the resistivity and anomalous transverse magnetoresistance or whether these observations can be understood in terms of the physics of the bulk.  Electronic structure calculations at reduced volume could be of great help in distinguishing between these possibilities.

\subsection*{Acknowledgments}
  Research at the University of California, San Diego, was supported by the U.\ S.\ Department of Energy (DOE) grant number DE-FG52-06NA26205.  Work at the University of Maryland was supported in part by the National Science Foundation MRSEC under Grant No. DMR-0520471.  N.\ P.\ B.\ is supported by CNAM.  Research at UAB was supported by NNSA-DOE grant DE-FG52-10NA29660.  Portions of this work were performed under LDRD.  JRJ and STW are supported by the Science Campaign at Lawrence Livermore National Laboratory. Lawrence Livermore National Laboratory is operated by Lawrence Livermore National Security, LLC, for the U.S. Department of Energy, National Nuclear Security Administration under Contract DE-AC52-07NA27344.  Portions of this work were performed at HPCAT (Sector 16), Advanced Photon Source (APS), Argonne National Laboratory. HPCAT is supported by CIW, CDAC, UNLV and LLNL through funding from DOE-NNSA, DOE-BES, and NSF. Use of the Advanced Photon Source, an Office of Science User Facility operated for the U.S. DOE Office of Science by Argonne National Laboratory, was supported by the U.S. DOE under Contract No. DE-AC02-06CH11357. Beamtime was provided through the General User Proposal program.


\begin{thebibliography}{10}

\bibitem{xia_2009_1}
Y.~Xia, D.~Qian, D.~Hsieh, L.~Wray, A.~Pal, H.~Lin, A.~Bansil, D.~Grauer, Y.~S.
  Hor, R.~J. Cava, and M.~Z. Hasan.
\newblock {\em Nature Phys.}, 5:398, 2009.

\bibitem{zhang_2009_1}
H.~Zhang, C.-X. Liu, X.-L. Qi, X.~Dai, Z.~Fang, and S.-C. Zhang.
\newblock {\em Nature Phys.}, 5:438, 2009.

\bibitem{butch_2010_1}
N.~P. Butch, K.~Kirshenbaum, P.~Syers, A.~B. Sushkov, G.~S. Jenkins, H.~D.
  Drew, and J.~Paglione.
\newblock {\em Phys. Rev. B}, 81:241301(R), 2010.

\bibitem{scifinder}
The CAS Database via SciFinder Scholar (accessed September 10, 2011).

\bibitem{fu_2008_1}
L.~Fu and C.~L. Kane.
\newblock {\em Phys. Rev. Lett.}, 100:0964007, 2008.

\bibitem{hasan_2010_1}
M.~Z. Hasan and C.~L. Kane.
\newblock {\em Rev. Mod. Phys.}, 82:3045, 2010.

\bibitem{zhang_2011}
J.~L. Zhang, S.~J. Zhang, H.~M. Weng, W.~Zhang, L.~X. Yang, Q.~Q. Liu, S.~M.
  Feng, X.~C. Wang, R.~C. Yu, L.~Z. Cao, L.~Wang, W.~G. Yang, H.~Z. Liu, W.~Y.
  Zhao, S.~C. Zhang, X.~Dai, Z.~Fang, and C.~Q. Jin.
\newblock {\em Proc. Nat. Acad. Sci}, 108:24, 2011.

\bibitem{hor_2010_1}
Y.~S. Hor, A.~J. Williams, J.~G. Checkelsky, P.~Roushan, J.~Seo, Q.~Xu, H.~W.
  Zandbergen, A.~Yazdani, N.~P. Ong, and R.~J. Cava.
\newblock {\em Phys. Rev. Lett.}, 104:057001, 2010.

\bibitem{vereshchagin_1965_1}
L.~F. Vereshchagin, E.~S. Itskevich, E.~Ya. Atabaeva, and S.~V. Popova.
\newblock {\em Sov. Phys. - Solid State}, 6:1763, 1965.

\bibitem{atabaeva_1973_1}
E.~Ya. Atabaeva, N.~A. Bendaliani, and S.~V. Popova.
\newblock {\em Fiz. Tverd. Tela}, 15:3508, 1973.

\bibitem{sushkov_2010}
A.~B. Sushkov, G.~S. Jenkins, D.~C. Schmadel, N.~P. Butch, J.~Paglione, and
  H.~D. Drew.
\newblock {\em Phys. Rev. B}, 82:125110, 2010.

\bibitem{jenkins_2010}
G.~S. Jenkins, A.~B. Sushkov, D.~C. Schmadel, N.~P. Butch, P.~Syers,
  J.~Paglione, and H.~D. Drew.
\newblock {\em Phys. Rev. B}, 82:125120, 2010.

\bibitem{cho_2011}
S.~Cho, N.~P. Butch, J.~Paglione, and M.~S. Fuhrer.
\newblock {\em Nano Lett.}, 2011.

\bibitem{fit2d}
A.~Hammersley, S.~Svensson, M.~Hanfland, A.~Fitch, and D.~Hausermann.
\newblock {\em High Press. Res.}, 14:235, 1996.

\bibitem{GSAS}
B.~H. Toby.
\newblock {\em J. Appl. Crystallogr.}, 34:210, 2001.

\bibitem{weir_2000_1}
S.~T. Weir, J.~Akella, C.~Aracne-Ruddle, Y.~K. Vohra, and S.~A. Catledge.
\newblock {\em Appl. Phys. Lett.}, 77:3400, 2000.

\bibitem{jackson_2006}
D.~D. Jackson, J.~R. Jeffries, W.~Qiu, J.~D. Griffith, S.~McCall, C.~Aracne,
  M.~Fluss, M.~B. Maple, S.~T. Weir, and Y.~K. Vohra.
\newblock {\em Phys. Rev. B}, 74:174401, 2006.

\bibitem{smith_1969_1}
T.~F. Smith, C.~W. Chu, and M.~B. Maple.
\newblock {\em Cryogenics}, 9:53, 1969.

\bibitem{wittig_1988_1}
B.~Bireckoven and J.~Wittig.
\newblock {\em J. Phys. E: Sci. Instrum.}, 21:841, 1988.

\bibitem{mao_1986_1}
H.~K. Mao, J.~Xu, and P.~M. Bell.
\newblock {\em J. Geophys. Res.}, 91:4673, 1986.

\bibitem{birch_1947_1}
F.~Birch.
\newblock {\em Phys. Rev.}, 71:809, 1947.

\bibitem{lundegaard_2005}
L.~F. Lundegaard, E.~Makovicky, T.~Boffa-Ballaran, and T.~Balic-Zunic.
\newblock {\em Phys. Chem. and Minerals}, 32:578, 2005.

\bibitem{hor_2009}
Y.~S. Hor, A.~Richardella, P.~Roushan, Y.~Xia, J.~G. Checkelsky, A.~Yazdani,
  M.~Z. Hasan, N.~P. Ong, and R.~J. Cava.
\newblock {\em Phys. Rev. B}, 79:195208, 2009.

\bibitem{liu_2011}
M.~Liu, C.~Z. Chang, Z.~Zhang, Y.~Zhang, W.~Ruan, K.~He, L.~Wang, X.~Chen,
  J.~F. Jia, S.~C. Zhang, Q.~K. Xue, X.~Ma, and Y.~Wang.
\newblock {\em Phys. Rev. B}, 83:165440, 2011.

\bibitem{cha_2010}
J.~J. Cha, J.~R. Williams, D.~Kong, S.~Meister, H.~Peng, A.~J. Bestwick,
  P.~Gallagher, D.~Goldhaber-Gordon, and Y.~Cui.
\newblock {\em Nano Lett.}, 10:1076, 2010.

\bibitem{paul_1961}
W.~Paul.
\newblock {\em J. of Appl. Phys.}, 32:2082, 1961.

\bibitem{alexander_1968}
M.~N. Alexander and D.~F. Holcomb.
\newblock {\em Rev. Mod. Phys.}, 40:815, 1968.

\bibitem{roth_1963}
H.~Roth, W.~D. Straub, W.~Bernard, and J.~E. {Mulhern Jr}.
\newblock {\em Phys. Rev. Lett.}, 11:328, 1963.

\bibitem{kawabata_1980}
A.~Kawabata.
\newblock {\em J. Phys. Soc. Japan}, 49:628, 1980.

\bibitem{kim_2011}
Y.~S. Kim, M.~Brahlek, N.~Bansal, E.~Edrey, G.~A. Kapilevich, K.~Iida,
  M.~Tanimura, Y.~Horibe, S.~W. Cheong, and S.~Oh.
\newblock {\em Phys. Rev. B}, 84:073109, 2011.

\bibitem{kim_2011_arxiv}
D. Kim, S. Cho, N.~P. Butch, P. Syers, K. Kirshenbaum,
  J. Paglione, and M.~S. Fuhrer.
\newblock {\em arXiv:/1105.1410}, 2011.

\bibitem{checkelsky_2009}
J.~G. Checkelsky, Y.~S. Hor, M.~H. Liu, D.~X. Qu, R.~J. Cava, and N.~P. Ong.
\newblock {\em Phys. Rev. Lett.}, 103:246601, 2009.

\bibitem{zhang_2011_crossover}
M.~L.~J. Zhang, C.~Z. Chang, Z.~Zhang, X.~Feng, K.~Li, K.~He, L.~Wang, X.~Chen,
  X.~Dai, Z.~Fang, et~al.
\newblock {\em arXiv:1103.3353}, 2011.

\bibitem{germanenko_1994}
A.~V. Germanenko and G.~M. Minkov.
\newblock {\em Phys. Stat. Sol. (b)}, 184:9, 1994.

\end{thebibliography}
\end{document}